# Modeling and analysis of an ultra-stable subluminal laser

Zifan Zhou[a,*], Joshua Yablon[a], Minchuan Zhou[b], Ye Wang[a], Alexander Heifetz[c], and M.S. Shahriar[a,b]

[a] Northwestern University, Department of Electrical Engineering and Computer Science, Evanston, IL 60208
[b] Northwestern University, Department of Physics and Astronomy, Evanston, IL 60208
[c] Argonne National Laboratory, Nuclear Engineering Division, Lemont, IL 60439

**Abstract**

We describe a subluminal laser which is extremely stable against perturbations. It makes use of a composite gain spectrum consisting of a broad background along with a narrow peak. The stability of the laser, defined as the change in frequency as a function of a change in the cavity length, is enhanced by a factor given by the group index, which can be as high as $10^5$ for experimentally realizable parameters. We also show that the fundamental linewidth of such a laser is expected to be smaller by the same factor. We first present an analysis where the gain profile is modeled as a superposition of two Lorentzian functions. We then present a numerical study based on a physical scheme for realizing the composite gain profile. In this scheme, the broad gain is produced by a high pressure buffer-gas loaded cell of rubidium vapor. The narrow gain is produced by using a Raman pump in a second rubidium vapor cell, where optical pumping is used to produce a Raman population inversion. We show close agreement between the idealized model and the explicit model. A subluminal laser of this type may prove to be useful for many applications.

[*] *Email address:* zifanzhou2012@u.northwestern.edu

# 1. Introduction

Recently, we have shown that the frequency of a superluminal ring laser (SRL) becomes highly sensitive to perturbations [1,2,3,4,5,6,7,8,9,10]. It has also been shown that the change in the resonant frequency of the so-called White Light Cavity [11,12,13,14,15], which is a passive version of the SRL and contains a critically tuned fast light medium, as a function of a change in the cavity length, is much larger than that for a conventional cavity [3,8]. Recent studies [2,3,16,17,18,19] have shown that the converse is also true. The resonant frequency of a cavity containing a slow light medium with a large group index changes very little as a function of a change in its cavity length, thus making it highly stable. In this paper, we describe the properties of the *active* version of such a slow light cavity: a subluminal laser (SLL). It can also be thought of as the slow-light counterpart of a superluminal laser.

Briefly, an SLL is a laser inside which the group velocity of light is much slower than the vacuum speed of light. The behavior of the SLL can be characterized by the group index, $n_g$. We show that the spectral sensitivity, defined as the change in lasing frequency as a function of a change in the cavity length, is reduced by a factor of $n_g$, the group index, which is defined as the ratio of the vacuum speed of light and the group velocity of light. Since values of $n_g$ as high as $10^5$ or more can be readily achieved, an SLL can be a super-stable laser.

Another interesting and potentially very important aspect of SLL is that its measured linewidth ($\gamma_{MEAS}$), under quantum noise and measurement bandwidth limits, which is given by the geometric mean of the Schawlow-Townes linewidth (STL) and the measurement bandwidth, can be substantially smaller than that of a conventional laser. As we show in detail later in this paper, this linewidth is expected to be smaller than that of a conventional laser by a factor of $n_g$. The large value of $n_g$ expected for an SLL would thus imply a very small value of $\gamma_{MEAS}$. A laser with such a small value of $\gamma_{MEAS}$, coupled with increased stability, may find important applications in many areas.

Of course, almost all lasers operate under conditions where $n_g > 1$. Thus, any such laser in principle can be called an SLL. What we describe here is a type of SLL where $n_g \gg 1$, with values as high $10^7$ for experimentally realizable parameters. In principle, an SLL with a significant group index can be realized by employing a gain medium with a very narrow gain spectrum. An example of such a laser is the Raman laser [20,21], in a Λ system where the gain spectrum can be very narrow (~MHz or smaller), with the linewidth being determined by a combination of the Rabi frequency of the Raman pump, the optical pumping (required for producing Raman population inversion between the low-lying states) rate, the collisional decay and dephasing rates for the two low-lying states, and the linewidth of the pump laser. In principle, such a Raman laser experiences the enhanced stability due to the slow-light effect. However, for experimentally accessible parameters, the amount of gain that can be produced in such a laser is very small, and decreases with decreasing linewidth of the gain spectrum, thus highly limiting the utility of such a laser. Finally, $\gamma_{MEAS}$ is inversely proportional to the root of the laser power. Thus, compared to an SLL with a higher power, $\gamma_{MEAS}$ for a Raman laser would be much larger. In this paper, we propose a model that produces an SLL with a stability enhancement factor ($n_g$) as large as $10^5$, without limiting its output power.

The rest of the paper is organized as follows. In section 2, we describe the proposed scheme for making an SLL, including the basic concept and the basic configuration. In section 3, the idealized analytical model is used for simulating the system. In section 4, the algorithm employed to solve the model using actual gain medium is explained in detail. In section 5, we explain and discuss the quantum noise and measurement bandwidth limited linewidth. A conclusion of the whole paper is presented in section 6.

## 2. Proposed scheme for making an SLL

In our proposed system for an SLL, as summarized in Fig. 1, the gain spectrum consists of two parts: a narrow gain peak added on top of a broad gain spectrum. We use Rb vapor as a specific example of a medium that can be used to realize this. The broad gain is provided by the same mechanism as is used for realizing the conventional DPAL (Diode Pumped Alkali Laser) [22,23,24,25,26,27]. A spectrally broad pump is applied on the $D_2$ line in a cell filled with a high pressure (~1 atm or more) of Ethane or $^4$He. Rapid relaxation (induced by the buffer gas) of atoms from the $P_{3/2}$ to the $P_{1/2}$ manifold thus leads to gain on the $D_1$ line, with a spectral width as much as 5 GHz for a buffer gas pressure of 1 atm, and much larger with higher buffer gas pressures. DPALs with output powers of close to 50 watt [25] have been reported in the literature. The gain peak is provided by adding to the cavity a second Rb cell, without any buffer. An optical pumping beam is applied to this cell to produce a Raman population inversion between the two hyperfine levels in the ground state ($|1\rangle$ and $|2\rangle$ in Fig. 1(b)). A piece of the linearly polarized output of the laser, at frequency $f_1$, is rotated in polarization by 90°, and shifted in frequency (by an acousto-optical modulator, AOM-1) by an amount that matches the ground state hyperfine splitting in Rb (~3.034 GHz for $^{85}$Rb, for example) to produce the Raman probe beam (at frequency $f_2$), which is applied to the second cell, by using a polarizing beam splitter (PBS), and then ejected from the lasing cavity by another PBS. Depending on the transition used for the optical pumping, this beam produces either a dip or a peak in the gain spectrum for the lasing field, with a width that can be as narrow as a few hundred KHz. When a gain dip is produced, the system becomes a superluminal laser under a carefully tuned set of parameters that produces $n_g$~0 [1]. When a gain peak is produced, the system becomes an SLL, with the enhancement in stability given by the value of $n_g$ corresponding to the gain peak, which can be as high as $10^5$.

It should be noted that in order for the lasing mode to experience a peak (or dip) in the gain, it is necessary for the frequency of the Raman probe beam to remain fixed at an absolute value. To accomplish this, a piece of the Raman probe beam is shifted in frequency by another AOM (AOM-2) to a value ($f_3$) that becomes resonant with an atomic transition (the $|2\rangle \leftrightarrow |3\rangle$ transition, for example). The beam at frequency $f_3$ is sent through a reference cell (e.g., a saturated absorption cell), and a feedback generated from the resonance observed in this cell is applied to the voltage controlled oscillator (VCO-1) that determines the frequency of AMO-1. Thus, if the frequency ($f_1$) of the laser moves (e.g., due to a change in the cavity length), the frequency output of VCO-1 is automatically adjusted to ensure that $f_2$ remains resonant, and thereby $f_1$ remains fixed, assuming that the frequency of the VCO (VCO-2) used for AOM-2 remains stable. If a frequency synthesizer is used instead of a VCO for AOM-2, it is possible to keep the frequency of AOM-2 (i.e., ($f_2$-$f_3$)) very stable. However, for Rb, Raman probe can be locked to a transition in $^{87}$Rb without shifting its frequency, which means that AOM-2 is not necessary in this approach. For example, the short term fractional frequency stability of an oven-controlled crystal oscillator can be $\sim 10^{-12}/\sqrt{\tau}$, where $\tau$ is the observation time (in unit of second). Long term stability can be provided by referencing this oscillator to an atomic clock, for example. We should note that a similar scheme can also be employed to increase the stability of a Raman laser, by locking a frequency-shifted piece of the Raman pump to an atomic transition.

## 3. Theoretical Model of an SLL using idealized gain media

We first consider a model of an SLL with idealized gain media, which can be analyzed easily through the semi-classical equation of motion for a single model laser [28]. Inside the laser cavity, the phase and the amplitude of the field, assumed to be in single mode, are described by the following equations [28]:

$$v + \dot{\varphi} = \Omega_c - \frac{\chi'}{2} v, \tag{1}$$

$$\dot{E} = -\frac{vE}{2Q} - \frac{\chi'' E}{2} v, \tag{2}$$

where $v$ is the lasing frequency; $\varphi$ and $E$ are the phase and the amplitude of the lasing field, respectively; $\Omega_c$ is the resonant frequency of the cavity without any medium, which is given by $2\pi mc/L$ where $L$ is the cavity length; $Q$ is the quality factor of the cavity; and $\chi'$ and $\chi''$ are the real and imaginary part of the susceptibility of the gain medium, respectively. Suppose $v_0$ is the frequency around which $\chi''$ is symmetric. We then define as $L_0$ the length of the cavity for which $\Omega_c = v_0$. In this model, while $L$ varies around $L_0$, $\Omega_c$ will derivate from $v_0$.

For convenience, we define the parameters: $\Delta \equiv \Omega_c - v_0, \delta \equiv v - v_0$. The derivatives $d\Delta/dL$ and $d\delta/dL$ characterize the resonant frequency shifts produced from a perturbation of $L$ in the empty and filled cavity, respectively. The ratio between the two derivatives, $R \equiv (d\delta/dL)/(d\Delta/dL)$, determines whether the amount of the frequency shift is enhanced ($R>1$) or suppressed ($R<1$) by the intracavity medium. To derive an explicit expression for $R$, we first solve Eq. 1 in steady state. After subtracting $v_0$ from both sides, differentiating with respect to $L$, and applying $dv = d\delta$, we get $d\delta/$ $dL + \chi'/2(d\delta/dL) + v/2(d\chi'/dL) = d\Delta/dL$ . By substituting $d\chi'/dL$ by $(d\chi'/d\delta)(d\delta/dL)$, $R$ can be expressed as:

$$R = 1 / \left(1 + \frac{\chi'}{2} + \frac{v}{2}\frac{d\chi'}{dv}\right). \tag{3}$$

Let us consider the case in which the cavity contains a medium with a narrow absorption as well as a medium with a broad gain. For simplicity we assume that the media overlap each other, and fill the whole cavity. This configuration creates a net gain profile with a dip in the center. The imaginary part of the susceptibility $\chi''$ can then be written as:

$$\chi'' = -\frac{G_e \Gamma_e^2}{2\Omega_e^2 + \Gamma_e^2 + 4(v-v_0)^2} - \frac{G_i \Gamma_i^2}{2\Omega_i^2 + \Gamma_i^2 + 4(v-v_0)^2}, \tag{4}$$

where the subscripts $e$ and $i$ refer to the envelop gain profile and the narrow gain profile, respectively; $G_k$ presents the gain parameter ($k=i, e$), which is defined as $G_k \equiv \hbar N_k \xi_k / \varepsilon_0$, where $N_k$ is the density of the quantum systems for the medium, $\xi_k$ is defined as $\xi_k \equiv \wp_k^2/(\hbar^2 \Gamma_k)$ using the Wigner-Weisskopf model [29] for spontaneous emission, and $\varepsilon_0$ is the permittivity of free space and $\wp_k$ is the dipole momentum. Here, $\Gamma_k$ presents the linewidth, $\Omega_k$ is the Rabi frequency, and $v$ is the lasing frequency. Then, applying the modified Kramers-Kronig relation [30,31,32], the real part of the susceptibility can be expressed as:

$$\chi' = \frac{2G_e(v-v_0)\Gamma_e}{2\Omega_e^2 + \Gamma_e^2 + 4(v-v_0)^2} + \frac{2G_i(v-v_0)\Gamma_i}{2\Omega_i^2 + \Gamma_i^2 + 4(v-v_0)^2}. \tag{5}$$

We fist consider a conventional, homogeneously broadened gain medium, i.e., $G_i = 0$. From Eq. 4 and Eq. 5, $\chi'$ and $\chi''$ are then simply related to each other as $\chi'/\chi'' = -2(v-v_0)/\Gamma_e$. Under steady state lasing condition, $\chi'' = -1/Q$, which leads to $\chi' = -2(v-v_0)/(Q\Gamma_e)$. As shown in this expression, $\chi'$ is a linear function of $v$, and antisymmetric around $v_0$. Noting that $Q = v_0/\Gamma_c$, where $\Gamma_c$ is the linewidth of the empty cavity, we find that $R = 1/n_g$, where $n_g = 1 + \Gamma_c/\Gamma_e$. Since $n_g > 1$, the sensitivity is suppressed by a conventional gain medium when compared to an empty cavity. In a typical laser, $\Gamma_c/\Gamma_e$ is very small, so that this reduction is rather insignificant.

For the case of a conventional laser medium discussed above, it was easy to determine the value of $\chi'$ due to the simple ratio between $\chi'$ and $\chi''$. In particular, this ratio does not depend on the laser intensity. However, for the SLL (i.e. $G_i \neq 0$), the two terms in $\chi''$ are highly dissimilar (as shown in Fig. 2), i.e., the amplitude and the linewidth of the narrow gain peak are significantly different from that of the envelop profile. Therefore, it is no longer possible to find a ratio between $\chi'$ and $\chi''$ that is independent of the laser intensity. Thus, in this case, it is necessary to determine first the manner in which the laser intensity depends on all the parameters, including $v$. We define $I \equiv |E|^2$ so that $\Omega_k^2 \equiv \Gamma_k I \xi_k$. Therefore, Eq. 4 and Eq. 5 become:

$$\chi'' = -\frac{G_e \Gamma_e^2}{2\Gamma_e \xi_e I + \Gamma_e^2 + 4(\nu-\nu_0)^2} - \frac{G_i \Gamma_i^2}{2\Gamma_i \xi_i I + \Gamma_i^2 + 4(\nu-\nu_0)^2}, \quad (6)$$

$$\chi' = \frac{2G_e(\nu-\nu_0)\Gamma_e}{2\Gamma_e \xi_e I + \Gamma_e^2 + 4(\nu-\nu_0)^2} + \frac{2G_i(\nu-\nu_0)\Gamma_i}{2\Gamma_i \xi_i I + \Gamma_i^2 + 4(\nu-\nu_0)^2}. \quad (7)$$

By setting Eq. 6 equal to $-1/Q$, we get $aI^2 + bI + c \equiv 0$, where a, b, and c are functions of various parameters. We keep the solution that is positive over the lasing bandwidth: $I = (-b + \sqrt{b^2 - 4ac})/(2a)$. Substituting this solution for $I$ to evaluate $\Omega_i^2$ and $\Omega_e^2$ in Eq. 7, we get an analytic expression for $\chi'$, which is plotted in Fig. 3. We note next that the steady state solution of Eq. 1 implies that $L = (2\pi mc/\nu)/(1 + \chi'/2)$. Using this expression, we can thus plot $L$ as a function of $\nu$, as shown in Fig. 4. All the parameters used for generating figures 2, 3, 4(a), and 5 are listed in Table 1.

**Table 1** Parameters used in the analytical model.

| Parameters | Value |
| --- | --- |
| $\Gamma_e$ | $2\pi \times 5 \times 10^8 \ s^{-1}$ |
| $\Gamma_i$ | $2\pi \times 10^2 \ s^{-1}$ |
| $\nu_0$ | $2\pi \times 3.8 \times 10^{14} \ s^{-1}$ |
| $N_e$ | $1 \times 10^6$ |
| $N_i$ | $1 \times 10^{14}$ |
| $Q$ | $1.5 \times 10^7$ |
| $m$ | 1282051 |
| $L_0$ | $0.99999978 \ m$ |
| $G_e$ | $6.667 \times 10^{-7}$ |
| $G_i$ | $6.667 \times 10^{-3}$ |

The condition used in generating Fig. 4(a) corresponds to a group index of $n_g(\nu = \nu_0) \sim 2.3 \times 10^5$, as noted earlier the factor by which frequency fluctuation is suppressed in an SLL is given by this value. As such, the conditions used here correspond to a highly stable SLL.

It should be noted, however, that when the cavity length changes sufficiently from its quiescent value of $L_0$, e.g.: $L - L_0 = 0.5 \times 10^{-5} m$, corresponding to the dashed vertical line in Fig. 4(a), there are multiple solutions of the laser frequency for a given length. In order to understand what this implies, we consider first a different set of parameters for which $n_g$ is much smaller $\left(n_g(v = v_0) \sim 160\right)$, and the corresponding plot of frequency versus length is shown in Fig. 4(b). This figure can be divided into three regions. In region 1 and 3, there exists only one solution. But in region 2, the laser equations have two or three solutions, i.e. for a specific cavity length, there are multiple potential lasing frequencies. However, the compound gain medium to be used for realizing the SLL is homogeneously broadened, as discussed in detail later in this paper. As such, mode competition is likely to ensure that only one of these solutions will survive in steady state. A detailed calculation involving nearly-degenerate multiple modes and the gain competition between these modes is needed to determine the exact behavior of the SLL in this region, and will be carried out in the near future, using the practical example we describe in the next section. Nonetheless, we can understand the behavior of the laser over a limited range of length changes. Assume, for example, that the cavity length is $L_0$ before the gain is activated. In this case, the SLL will lase at $v = v_0$, corresponding to the point O in Fig. 4(b). If $\Delta L$ now stays small enough so that the SLL stays within zone 1, the SLL will move along the line AOA'. If $\Delta L$ is large enough, but still smaller than $|OB| = |OB'|$, the SLL will stay on the BAOA'B' line, since the gain in the two other possible modes (within zone 2) will be highly suppressed, given the fact that the SLL is already lasing in the primary mode. Similarly, if $\Delta L$ is large enough so that the SLL moves to zone 3, it will jump to the unique frequency corresponding to that zone. The only concern would be when the value of $\Delta L$ is such that the SLL reaches point C or C'. In that case, it is not clear whether its frequency will follow the CA line or the CD line when $\Delta L$ is reduced. As stated above, this question can only be answered by a study of the temporal dynamics of the cavity field. Thus, if the SLL is operated in such a manner that $\Delta L$ remains less than $|OC| = |OC'|$, it will operate in the stable mode, which will experience the enhanced stability in frequency. In what follows, we will refer to this range as the primary mode zone, and the stable mode as the primary mode.

In Fig. 5(a) we plot the output intensity of the laser as a function of the cavity length, for the parameters used in producing Fig. 4(b). As can be seen, the intensity has a unique value for some ranges of the cavity length, but has multiple possible values over the ranges for which the frequency also has multiple values (see Fig. 4(b)). Again, for our current model, it is not clear as to what the actual value of the intensity would be, and this issue can only be resolved by studying the temporal dynamics, as noted above.

In Fig. 5(b), we show the output intensity as a function of the laser frequency. This plot is generated by combining the information presented in Fig. 4(b) and Fig. 5(a). As can be seen, there is a one-to-one correspondence between the intensity and the frequency, for all values of the cavity length. In Fig. 6(a), we show the laser frequency versus the cavity length for the primary mode, corresponding to the parameters used in Fig. 4(a). The resulting enhancement in stability is shown in Fig. 6(b). As can be seen, the stability of the SLL can be $\sim 10^5$ times higher than that of a conventional laser when operating near the center frequency.

## 4. Theoretical model of an actual SLL

The gain spectrum we have considered so far is, of course, only an ideal one. The behavior of a real system would certainly deviate from such a model. Here, we consider an atomic system that can be tailored to produce a gain spectrum that is very similar to the ideal one presented above, and determine the expected behavior of an SLL potentially realizable using such a system.

The broad gain profile can be realized using an alkali vapor cell, containing both naturally occurring isotopes of Rb ($^{87}$Rb and $^{85}$Rb), in the presence of a high pressure buffer gas of Ethane. Such a gain medium has recently been used to make the so called DPAL (diode pumped alkali laser). It has also been used recently by us to demonstrate a high-speed optical modulator [33]. The narrow gain profile can be realized using a separate vapor cell, placed in series, containing only $^{85}$Rb, configured for Raman gain.

In this case we find the dispersion and absorption of the media by solving the density matrix equations in steady state. Specifically, we employ the Liouville equation, which describes the evolution of the density matrix:

$$\frac{\partial}{\partial t}\tilde{\rho} = -\frac{i}{\hbar}\left[\widetilde{\tilde{H}},\tilde{\rho}\right] + \frac{\partial \tilde{\rho}}{\partial t}_{source} + \frac{\partial \tilde{\rho}}{\partial t}_{dephasing}, \tag{8}$$

where, $\tilde{\rho}$ is the density matrix in the rotating wave basis, $\widetilde{\tilde{H}}$ is the modified time independent Hamiltonian under rotating wave approximations (RWA) augmented by adding complex terms to its diagonal elements in order to represent decays of atomic levels, $\tilde{\rho}_{source}$ represents the influx of atoms into a state due to decay from another state, and $\tilde{\rho}_{dephasing}$ accounts for the dephasing between states induced by the buffer gas [34]. Given the different configuration and parameters used for the two cells, it is necessary to solve the Liouville equation separately for each cell.

*4.1 Modeling the broad gain system*

For the cell that produces the DPAL-type gain, we use a laser beam tuned to the D2 transition of Rb as the optical pump. The specific parameters, including the frequency of this pump, are specified later in this section. This pump couples atoms from both hyperfine levels in the $^5S_{1/2}$ states to the $^5P_{3/2}$ manifold. The presence of a high pressure buffer gas (~0.5 atm of Ethane) causes rapid dephasing of the coherence corresponding to the $^5S_{1/2}$, F=1 - $^5P_{3/2}$ transition as well as the coherence corresponding to the $^5S_{1/2}$, F=2 - $^5P_{3/2}$ transition, thus producing a homogeneous broadening of each of these transitions by ~7GHz (for 0.5 atm pressure of Ethane). Furthermore, the buffer gas cause the atom in the $^5P_{3/2}$ manifold to decay rapidly to the $^5P_{1/2}$ manifold, at a rate that is much faster than the radiative decay rate for the atoms in the $^5P_{1/2}$ manifold. As a result, a population inversion is created between the $^5P_{1/2}$ manifold and the $^5S_{1/2}$ manifold. This inversion provides the broad band gain profile.

The homogeneous broadening induced by the buffer gas is significantly larger than the hyperfine splittings within the $^5P_{1/2}$ and the $^5P_{3/2}$ manifolds. However, it is of the order of the hyperfine splitting in the $^5S_{1/2}$ manifold. Thus, we treat the $^5P_{1/2}$ and $^5P_{3/2}$ manifolds as single energy levels, but keep track of the two hyperfine levels inside the $^5S_{1/2}$ manifolds separately. Therefore, we have a four level system, as shown in Fig. 7(a) shows the system for $^{85}$Rb. Here

|1⟩ is the $^5S_{1/2}$, F=2 state, |2⟩ is the $^5S_{1/2}$, F=3 state, |3⟩ is the $^5P_{1/2}$ manifold, and |4⟩ is the $^5P_{3/2}$ manifold. The corresponding system for $^{87}$Rb is shown in Fig. 7(b), where |1⟩ is the $^5S_{1/2}$, F=1 state, |2⟩ is the $^5S_{1/2}$, F=2 state, |3⟩ is the $^5P_{1/2}$ manifold, and |4⟩ is the $^5P_{3/2}$ manifold. In our model, we consider a cell that contains a natural mixture (72.16% of $^{85}$Rb, 27.84% of $^{87}$Rb) of both stable isotopes of Rubidium. We compute the gain produced by each isotope separately, and add the weighted sum to find the net gain. In doing so, we keep track of the differences in the absolute frequencies of the transitions in these two isotopes. For each isotope of Rb, we assume that the pump beam excites both the |1⟩ − |4⟩ and the |2⟩ − |4⟩ transitions. We assume the transition strengths (i.e. the Rabi frequencies) to be the same, for both transitions, for simplicity. A more accurate calculation would require taking into account the hyperfine levels within the $^5P_{3/2}$ manifold, as well as the Zeeman sublevels within each hyperfine states, and the various dipole-moment matrix elements among them [34]. Such a calculation will be carried out and reported in the near future. Similarly, for each isotope of Rb, we assume that the probe beam excites both the |1⟩ − |3⟩ and the |2⟩ − |3⟩ transitions, with equal transition strengths.

The relevant decay and dephasing rates are illustrated schematically in Fig. 8. In Fig. 8(a), we show only the rates of population decays, due to radiative as well as non-radiative (i.e. collisional) processes. Specifically, $\Gamma_{r3}$ and $\Gamma_{r4}$ are the radiative population decay rates which are the inverses of the radiative lifetimes 27.7 ns and 26.24 ns [35,36], respectively. Thus, $\Gamma_{r3} = 36.1 \times 10^6\ sec^{-1}$, and $\Gamma_{r4} = 38.1 \times 10^6\ sec^{-1}$. For simplicity, we again assume that the decay rate of atoms from |4⟩ to |1⟩ is the same as that of atoms from |4⟩ to |2⟩, given by $\Gamma_{r4}/2$. As noted earlier, use of a more accurate branching ratio would require taking into account the dipole-moment matrix elements coupling all the Zeeman sublevels. A detailed calculation of this type would be carried out and reported in the near future. Similarly, we assume that the decay rates to |1⟩ and |2⟩ from state |3⟩ are the same, given by $\Gamma_{r3}/2$. The collisional decay rate, $\Gamma_{43}$, at a temperature of ~100 °C, is known [24] to be ~$1.41 \times 10^7\ sec^{-1}/Torr$. For a buffer gas pressure of 0.5 atm (~380 Torr) assumed here, we thus have $\Gamma_{43} \simeq 5.36 \times 10^9\ sec^{-1}$. The value of $\Gamma_{34}$ is determined by assuming thermal equilibrium, obeying the Boltzmann relation. Give that the number of Zeeman sublevels in the $^5P_{3/2}$ manifold is twice as large as the number of Zeeman sublevels in the $^5P_{1/2}$ manifold, for both isotopes, we have:

$$\frac{\Gamma_{34}}{\Gamma_{43}} = 2exp\left(-\frac{\Delta E_{43}}{k_B T}\right), \tag{9}$$

where $\Delta E_{43}$ is the energy difference between states $|4\rangle$ and $|3\rangle$, T is the temperature in Kelvin, and $k_B$ is the Boltzmann constant. Given that $\Delta E_{43}$ is approximately the same in both $^{85}$Rb and $^{87}$Rb, we get that $\Gamma_{34} \simeq 4.27 \times 10^9 \, sec^{-1}$. Similarly, the decay rates $\Gamma_{12}$ and $\Gamma_{21}$ are related to each other as:

$$\frac{\Gamma_{12}}{\Gamma_{21}} = \frac{5}{3}exp\left(-\frac{\Delta E_{21}}{k_B T}\right), \tag{10}$$

where, for both isotopes, the ratio of the number of Zeeman sublevels in levels $|2\rangle$ and $|1\rangle$ is 5/3, and $\Delta E_{21}$ is the energy difference between levels $|2\rangle$ and $|1\rangle$. Since $k_B T \gg \Delta E_{21}$ for both isotopes, we get that $\Gamma_{12} \simeq 1.67\Gamma_{21}$. In the absence of buffer gas, the value of $(\Gamma_{21}/2\pi)$ is $\sim 1 \, MHz$. The exact effect of the high pressure buffer gas (such as Ethane used to produce DPAL type gain) on this rate has, to the best of our knowledge, not yet been studied, experimentally or theoretically. We assume this effect to be small, and choose $\Gamma_{21} \simeq 2\pi \times 10^6 \, sec^{-1}$, so that $\Gamma_{12} \simeq 2\pi \times 1.6 \times 10^6 \, sec^{-1}$. These decay terms also contribute to dephasing of the coherence among these states. Specifically, if a state $|i\rangle$ has a net population decay rate of $\gamma_i$, and state $|j\rangle$ has a net population decay rate of $\gamma_j$, then the $\tilde{\rho}_{ij}$ coherence decays at the rate of $\frac{1}{2}(\gamma_i + \gamma_j)$. As mentioned above, the population decay rates, as well as the dephasing due to these decays, are taken into account by adding a term $-i\hbar\gamma_j/2$ to the j-th diagonal element of the Hamiltonian. The influxes of atoms into various states, on the other hand, are taken into account by the second term in Eq. 8.

The coherence between two states can undergo additional dephasing due to collisions with the high pressure buffer gas. These buffer-gas-induced (BGI) dephasing rates are illustrated in Fig. 8(b), and are accounted for by the third term in Eq. 8. The BGI dephasing rates of $\tilde{\rho}_{13}$, $\tilde{\rho}_{23}$, $\tilde{\rho}_{14}$, and $\tilde{\rho}_{24}$ are nearly equal to one another, denoted as $\Gamma_d$. The value of $\Gamma_d$ is known [25,37] to be $\sim 2\pi \times 2 \times 10^7 \, sec^{-1}/Torr$, when Ethane is used as a buffer gas. Thus, for a buffer gas pressure of 0.5 atm ( $\sim$380 Torr) assumed here, we get $\Gamma_d \simeq 2\pi \times 7.6 \times 10^9 \, sec^{-1}$. On the other hand, the BGI dephasing rates of $\tilde{\rho}_{34}$ and $\tilde{\rho}_{12}$ for high pressure buffer gases have not been investigated theoretically or experimentally. However, based on the physical processes involved in producing the BGI dephasing of the optical coherences (i.e. $\tilde{\rho}_{13}$, $\tilde{\rho}_{23}$, $\tilde{\rho}_{14}$, and $\tilde{\rho}_{24}$), it is reasonable to assume that these rates are comparable to $\Gamma_d$. Specifically, we write the BGI dephasing rate of $\tilde{\rho}_{12}$ ($\tilde{\rho}_{34}$) as $\alpha\Gamma_d$ ($\beta\Gamma_d$), where $\alpha$ and $\beta$ are parameters with values of the order of unity. Here, we calculated the DPAL gain with different values of $\alpha$ and $\beta$. The maximum DPAL gain versus $\alpha$ and $\beta$ are shown in Fig. 9. In Fig. 9(a), we vary the value of $\alpha$, and both the maximum value and the shape of DPAL gain profile almost remain the same. Similarly, as shown in Fig. 9(b), $\beta$ also does not influence the DPAL gain much. Since the amount of $\tilde{\rho}_{12}$ or $\tilde{\rho}_{34}$ coherence produced is expected to be very small, the gain experienced by the probe is expected to depend very weakly on the actual values of $\alpha$ and $\beta$, as we will show later. Based on these findings, we will use $\alpha = \beta \simeq 1$ for our model.

The modified time independent Hamiltonian under RWA augmented by adding complex terms to its diagonal elements of $^{85}$Rb and $^{87}$Rb can be written as:

$$\frac{2\widetilde{H}_k}{\hbar} = (-i\Gamma_{12})|1\rangle\langle 1| + (2\omega_k - i\Gamma_{21})|2\rangle\langle 2| + [-2\delta_s - i(\Gamma_{r3} + \Gamma_{34})]|3\rangle\langle 3| + [-2\delta_p - i(\Gamma_{r4} + \Gamma_{43})]|4\rangle\langle 4| + \{[\Omega_s|1\rangle\langle 3| + \Omega_p|1\rangle\langle 4| + \Omega_s|2\rangle\langle 3| + \Omega_p|2\rangle\langle 4|] + h.c.\}, \quad (11)$$

where the subscript $k$ refers to $^{85}$Rb or $^{87}$Rb; $\Omega_s$ and $\Omega_p$ are the Rabi frequencies of the signal field and the optical pump field, respectively; $\delta_s$ is the detuning of the signal field with respect to the $|1\rangle - |3\rangle$ transition; $\delta_p$ is the detuning of the pump field with respect to the $|1\rangle - |4\rangle$ transition; and $\omega_k$ is the energy difference between levels $|2\rangle$ and $|1\rangle$ with the different values for the two isotopes. The influx of atoms into various states is:

$$\frac{\partial \tilde{\rho}}{\partial t}\bigg|_{source} = (\Gamma_{21}\tilde{\rho}_{22} + \Gamma_{r3}\tilde{\rho}_{33}/2 + \Gamma_{r4}\tilde{\rho}_{44}/2)|1\rangle\langle 1| + (\Gamma_{12}\tilde{\rho}_{11} + \Gamma_{r3}\tilde{\rho}_{33}/2 + \Gamma_{r4}\tilde{\rho}_{44}/2)|2\rangle\langle 2| + \Gamma_{43}\tilde{\rho}_{44}|3\rangle\langle 3| + \Gamma_{34}\tilde{\rho}_{33}|4\rangle\langle 4|, \quad (12)$$

And the matrix which describes the BGI dephasing is:

$$\frac{\partial \tilde{\rho}}{\partial t}\bigg|_{dephasing} = -[\alpha\Gamma_d\tilde{\rho}_{12}|1\rangle\langle 2| + \Gamma_d\tilde{\rho}_{13}|1\rangle\langle 3| + \Gamma_d\tilde{\rho}_{14}|1\rangle\langle 4| + \Gamma_d\tilde{\rho}_{23}|2\rangle\langle 3| + \Gamma_d\tilde{\rho}_{24}|2\rangle\langle 4| + \beta\Gamma_d\tilde{\rho}_{24}|3\rangle\langle 4| + h.c.]. \quad (13)$$

A typical gain spectrum of the DPAL with $^{85}$Rb, $^{87}$Rb, and Ethane buffer gas is shown in Fig. 10. The parameters used in the calculation, such as cell length, temperature, buffer gas pressure, etc., are listed in Table 2.

**Table 2** Parameters used for generating gain spectrum of the DPAL.

| Parameters | Value |
| --- | --- |
| Total cavity length | $0.72\ m$ |
| Gain cell length | $0.1\ m$ |
| Gain cell temperature | $100\ °C$ |
| D1 transition wavelength | $795\ nm$ |
| D1 transition life time | $27.7\ nsec$ |
| D2 transition wavelength | $780\ nm$ |
| D2 transition life time | $26.24\ nsec$ |
| Radiative decay rate from $|2\rangle$ to $|1\rangle$ | $2\pi \times 1 \times 10^6\ sec^{-1}$ |
| Radiative decay rate from $|1\rangle$ to $|2\rangle$ | $2\pi \times 1.67 \times 10^6\ sec^{-1}$ |
| Ethane buffer gas pressure | $380\ Torr$ |
| Collisional decay rate from $|4\rangle$ to $|3\rangle$ | $2\pi \times 8.536 \times 10^8\ sec^{-1}$ |
| Collisional decay rate from $|3\rangle$ to $|4\rangle$ | $2\pi \times 6.796 \times 10^8\ sec^{-1}$ |
| BGI dephasing rate between $|2\rangle$ and $|1\rangle$ | $2\pi \times 20 \times 10^6\ (sec \cdot Torr)^{-1}$ |
| BGI dephasing rate between $|4\rangle$ and $|3\rangle$ | $2\pi \times 20 \times 10^6\ (sec \cdot Torr)^{-1}$ |
| BGI dephasing rate between any other two levels | $2\pi \times 20 \times 10^6\ (sec \cdot Torr)^{-1}$ |
| Frequency difference between $|1\rangle$ and $|2\rangle$ ($^{85}$Rb)[35] | $2\pi \times 3.034 \times 10^9\ sec^{-1}$ |
| Frequency difference between $|1\rangle$ and $|2\rangle$ ($^{87}$Rb)[36] | $2\pi \times 6.835 \times 10^9\ sec^{-1}$ |
| Frequency difference between $|1\rangle$ of $^{85}$Rb and $|1\rangle$ of $^{87}$Rb | $2\pi \times 2.501 \times 10^9\ sec^{-1}$ |
| Saturated intensity | $120\ W/m^2$ |

*4.2 Modeling the narrow gain system*

As stated above, the narrow gain profile is realized by inserting a separate vapor cell (Raman cell) inside the cavity to produce Raman Gain. This vapor cell is filled with pure $^{85}$Rb without any buffer gas. The basic scheme for this system is shown in Fig. 11(a). Briefly, an optical pumping beam couples state $|1\rangle$ to state $|4\rangle$. This produces a population inversion among levels $|1\rangle$ and $|2\rangle$. A Raman pump is now applied on the $|2\rangle - |3\rangle$ transition, but detuned significantly above resonance. Under this condition, the probe beam (which is now assumed to excite the $|1\rangle - |3\rangle$ transition only) experiences a narrow-band Raman gain centered around the two photon resonance condition (i.e. the frequency difference between the probe and the Raman pump matches the energy separation between $|1\rangle$ and $|2\rangle$.

In Fig. 11(b), we show the effective 3-level system and the various decay rates relevant to this excitation. For simplicity, we assume that the radiative decay rates from $|3\rangle$ to $|2\rangle$ and that from $|3\rangle$ to $|1\rangle$ are equal ($\Gamma_{r3}/2$). The decay rate from $|2\rangle$ to $|1\rangle$ is $\Gamma_{21}$, and that from $|1\rangle$ to $|2\rangle$ is $\Gamma'_{12} = \Gamma_{12} + \Gamma_{op}$. Here, $\Gamma_{12}$ and $\Gamma_{21}$ are collisional decay rates that are related to each other by Eq. 10, which in turn implies that $\Gamma_{12} \simeq 1.67\Gamma_{21}$. The additional decay from $|1\rangle$ to $|2\rangle$ at the rate of $\Gamma_{op}$ accounts for the effect of the optical pumping. The density matrix equation of evolution for this system can now be written as:

$$\frac{\partial}{\partial t}\tilde{\rho} = -\frac{i}{\hbar}\left[\widetilde{H}_R, \tilde{\rho}\right] + \frac{\partial \tilde{\rho}}{\partial t}_{source-R}. \tag{14}$$

Here,

$$2\widetilde{H}_R/\hbar = (-i\Gamma'_{12})|1\rangle\langle 1| + (-2\delta_s + 2\delta_{Rp} - i\Gamma_{21})|2\rangle\langle 2| + (-2\delta_s - i\Gamma_{r3})|3\rangle\langle 3| + (\Omega_s|1\rangle\langle 3| + \Omega_{Rp}|2\rangle\langle 3| + h.c.), \tag{15}$$

and

$$\frac{\partial \tilde{\rho}}{\partial t}_{source-R} = (\Gamma_{21}\tilde{\rho}_{22} + \Gamma_{r3}\tilde{\rho}_{33}/2)|1\rangle\langle 1| + (\Gamma'_{12}\tilde{\rho}_{11} + \Gamma_{r3}\tilde{\rho}_{33}/2)|2\rangle\langle 2|. \tag{16}$$

A typical gain spectrum of the Raman cell is shown in Fig. 12. The parameters used in the calculation are listed in Table 3.

**Table 3** Parameters used for generating gain spectrum of the medium in Raman cell.

| Parameters | Value |
|---|---|
| Total cavity length | $0.72\ m$ |
| Raman cell length | $0.1\ m$ |
| Raman cell temperature | $100\ °C$ |
| D1 transition wavelength | $795\ nm$ |
| D1 transition life time | $27.7\ nsec$ |
| Radiative decay rate from $|2\rangle$ to $|1\rangle$ | $2\pi \times 1 \times 10^6\ sec^{-1}$ |
| Radiative decay rate from $|1\rangle$ to $|2\rangle$ | $2\pi \times 1.67 \times 10^6\ sec^{-1}$ |
| Effective decay rate from $|1\rangle$ to $|2\rangle$ | $2\pi \times 1.67 \times 10^7\ sec^{-1}$ |
| Frequency difference between $|1\rangle$ and $|2\rangle$ ($^{85}$Rb)[35] | $2\pi \times 3.034 \times 10^9\ sec^{-1}$ |
| Raman pump detuning | $-2\pi \times 5 \times 10^9\ sec^{-1}$ |
| Saturated intensity | $120\ W/m^2$ |

*4.3 Combined effective susceptibility of the SLL*

After solving the density matrix equation of evolution for the two media in steady state, we can calculate the susceptibilities of the two media using the expression [34]:

$$\chi_D = \frac{\hbar c n_D}{\mathcal{I}_{sat}\Omega_s}\left(\frac{\Gamma_{r3}}{2}\right)^2 (\tilde{\rho}_{31} + \tilde{\rho}_{32}), \qquad (17)$$

$$\chi_R = \frac{\hbar c n_R}{\mathcal{I}_{sat}\Omega_s}\left(\frac{\Gamma_{r3}}{2}\right)^2 \tilde{\rho}_{31}, \qquad (18)$$

where $n_D$ and $n_R$ are the number density of the atoms in the DPAL and Raman media, respectively, and $\mathcal{I}_{sat}$ (in unit of Watts/m²) is the saturation intensity which corresponds to a field that produce $\Omega_s = \Omega_{sat} \equiv \Gamma_{r3}/2$. We assume this saturated intensity to be twice as big as that for the strongest transition in the D2 manifold. Then we can write the effective susceptibility of the media in the SLL as:

$$\chi_{eff} = \frac{L_D}{L}(0.72\chi_{D85} + 0.28\chi_{D87}) + \frac{L_R}{L}\chi_R, \qquad (19)$$

where $L$ is the cavity length, and $L_D$ and $L_R$ are the DPAL cell length and the Raman cell length, respectively. The real and imaginary parts of $\chi_{eff}$ must satisfy Eq. 1 and Eq. 2 in steady state. Thus, Eq. 19 and the steady state version of Eq. 1 and Eq. 2 have to be solved simultaneously, in a self-consistent manner. This requires the use of an iterative algorithm to find the value of $\delta_s$ and $\Omega_s$ (which in turn yields the frequency $\nu$ and the field amplitude $E$) that satisfy these equations; which is described next.

*4.4 Iterative algorithm for finding the frequency and the amplitude of the SLL*

The flow chart shown in Fig. 13 illustrates how the algorithm works. For a certain cavity length, this algorithm is able to find the frequency $\nu$ and the intensity $\mathcal{I}$ of the laser field in steady state. The frequency and the intensity of the field can be calculated from $\delta_s$ and $\Omega_s$ using the following equations:

$$\nu = \nu_0 - \delta_{Rp} + \delta_s, \qquad (20)$$

$$\mathcal{I} = \mathcal{I}_{sat} \left(\frac{\Omega_s}{\Omega_{sat}}\right)^2, \qquad (21)$$

where $\nu_0, \delta_{Rp}, \delta_s$ are as defined in Fig. 11.

The algorithm starts by assuming a pair of values of $\delta_s$ and $\Omega_s$. These yield a value of $\chi_{eff}$, by evaluating Eq. 19. This value of $\chi_{eff}$ (real and imaginary parts) is then fed into the steady state forms of the laser equations (Eq. 1 and Eq. 2) to find the values of $\delta'_s$ and $\Omega'_s$. These values are then compared with the values of $\delta_s$ and $\Omega_s$. If the differences $|\delta_s - \delta'_s|$ and $|\Omega_s - \Omega'_s|$ are each below chosen threshold values, the algorithm stops. If not, then either $\delta'_s$ or $\Omega'_s$ or both are increased of decreased by a suitable step size, and the loop is repeated, until a convergence is found. We note that, under certain conditions, there exist more than one set of solutions (i.e. combination of laser frequency and intensity) for a given set of parameters and cavity length. In these cases, we have found that an alternative form of the algorithm is more convenient to use. In this form, we fix the value of the laser frequency, and then iterate to find the values of the laser intensity and the cavity length that satisfy the relevant equations.

To explore the relationship between the lasing frequency and cavity length variations, we use this algorithm to find the lasing frequency in steady state for different cavity lengths. Then the stability improvement of the SLL can be calculated by comparing the derivative of the frequency shift as a function of the cavity length with that of a conventional laser.

*4.5 Results*

In Fig. 14(a), we show the output frequency as the cavity length is varied. Here, $L_0$ is a reference cavity length for which one of the cavity modes matches the value of $\nu_0$ (note that the free spectral range is much larger than the width of the Raman gain). As can be seen, the shift in the laser frequency is small for a significant spread in $\Delta L$. Furthermore, there is a range of cavity lengths for which multiple frequencies satisfy the equations. This is similar to the behavior described in our analytical model earlier. In Fig. 14(b), we again show a plot of the laser frequency as a function of the cavity length, but for a different set of parameters. As can be seen, there is a range around $L = L_0$ for which the frequency has a unique value, with a small slope, corresponding to reduced stability. The shaded area corresponds to the range of cavity length for which the laser frequency is not unique, same as what we observed earlier in Fig. 4(b). As we had noted earlier, a detailed calculation involving nearly-degenerate multiple modes and the gain competition between these modes is needed to determine the exact behavior of the SLL in this region. Such a calculation entails the use of an interaction Hamiltonian that cannot be rendered time independent via rotating wave transformation. As such, many higher orders terms in the solution of the density matrix equations have to be computed, in a manner akin to what is done in ref. 38, although in a different context. This analysis will be carried out and reported in the near future. Outside the shaded area, the frequency is again unique. However, the asymptotic slope is now much larger than that around $L = L_0$, corresponding to the normal sensitivity.

In Fig. 15(a), we show the laser power as a function of the cavity length, for same parameters we used in producing Fig. 14(b). As can be seen, the intensity has a unique value for some ranges of the cavity length, but has multiple possible values over the ranges for which the frequency also has multiple values (see Fig. 14(b)). Again, this behavior is similar to what we presented earlier for the analytical model. In Fig. 15(b), we show the output intensity as a function of the laser frequency. This plot is generated by combining the information presented in Fig. 14(b) and Fig. 15(a). As can be seen, there is a one-to-one correspondence between the intensity and the frequency, for all values of the cavity length.

In Fig. 16(a), we show the laser frequency as a function of the cavity length, for the primary mode, corresponding to the parameters used in Fig. 14(a). The resulting enhancement in stability is shown in Fig. 16(b). The sensitivity of the laser is reduced by a factor of $10^5$ when the lasing frequency is at $\nu = \nu_0$, corresponding to $L = L_0$. Finally, Fig. 17 shows the real part of the effective dispersion, $\chi_{eff}$, as a function of the laser frequency. As can be seen, the effective dispersion has qualitatively the same shape as that of the analytical approach.

**Table 4** Parameters used in the numerical model.

| Parameters | Value |
|---|---|
| Total cavity length | $0.72\ m$ |
| Gain cell length | $0.1\ m$ |
| Raman cell length | $0.1\ m$ |
| Gain cell temperature | $100\ °C$ |
| Raman cell temperature | $100\ °C$ |
| D1 transition wavelength | $795\ nm$ |
| D1 transition life time | $27.7\ nsec$ |
| D2 transition wavelength | $780\ nm$ |
| D2 transition life time | $26.24\ nsec$ |
| Radiative decay rate from $|2\rangle$ to $|1\rangle$ | $2\pi \times 1 \times 10^6\ sec^{-1}$ |
| Radiative decay rate from $|1\rangle$ to $|2\rangle$ | $2\pi \times 1.6 \times 10^6\ sec^{-1}$ |
| Effective decay rate from $|1\rangle$ to $|2\rangle$ | $2\pi \times 1.6 \times 10^7\ sec^{-1}$ |
| Ethane buffer gas pressure | $380\ Torr\ (0.5\ atm)$ |
| Collisional decay rate from $|4\rangle$ to $|3\rangle$ | $2\pi \times 8.536 \times 10^8\ sec^{-1}$ |
| Collisional decay rate from $|3\rangle$ to $|4\rangle$ | $2\pi \times 6.796 \times 10^8\ sec^{-1}$ |
| BGI dephasing rate between $|2\rangle$ and $|1\rangle$ | $2\pi \times 20 \times 10^6\ (sec \cdot Torr)^{-1}$ |

| | |
|---|---|
| BGI dephasing rate between $|4\rangle$ and $|3\rangle$ | $2\pi \times 20 \times 10^6\ (sec \cdot Torr)^{-1}$ |
| BGI dephasing rate between any other two levels | $2\pi \times 20 \times 10^6\ (sec \cdot Torr)^{-1}$ |
| Frequency difference between $|1\rangle$ and $|2\rangle$ ($^{85}$Rb)[35] | $2\pi \times 3.034 \times 10^9\ sec^{-1}$ |
| Frequency difference between $|1\rangle$ and $|2\rangle$ ($^{87}$Rb)[36] | $2\pi \times 6.835 \times 10^9\ sec^{-1}$ |
| Frequency difference between $|1\rangle$ of $^{85}$Rb and $|1\rangle$ of $^{87}$Rb | $2\pi \times 2.501 \times 10^9\ sec^{-1}$ |
| Rabi frequency of the DPAL pump | $2\pi \times 1.516 \times 10^9\ sec^{-1}$ |
| Rabi frequency of the Raman pump | $2\pi \times 2.873 \times 10^8\ sec^{-1}$ |
| Raman pump detuning | $-2\pi \times 1.6 \times 10^9\ sec^{-1}$ |
| Saturated intensity | $120\ W/m^2$ |

## 5. Quantum Noise Limited Linewidth of the SLL

As mentioned earlier, an interesting and potentially very important aspect of the SLL is its quantum noise limited linewidth. For a conventional laser with a very flat gain spectrum, the STL can be expressed simply as [39]:

$$\gamma_{STL} = \frac{\hbar\omega}{2P_{out}\tau_c^2}, \tag{22}$$

where ω is the laser frequency, $P_{out}$ is the output power, and $\tau_c$ is the lifetime of the empty cavity, which can also be expressed simply as the invers of the empty cavity linewidth. In the presence of a non-negligible normal dispersion (but not for anomalous dispersion [40]), this expression is modified significantly [41] as follows:

$$\gamma_{STL} = \frac{\hbar\omega}{2P_{out}(n_g\tau_c)^2}(1+|\alpha|^2), \tag{23}$$

where, $n_g$ is the group index, and $|\alpha| \equiv \left(\frac{\partial n'}{\partial |E|}\right) / \left(\frac{\partial n''}{\partial |E|}\right)$, where $n'$ and $n''$ are, respectively, the real and imaginary parts of the complex refractive index:

$$n_c \equiv n' + in'' = \sqrt{1+\chi} \approx \left(1 + \frac{1}{2}\chi'\right) + i\frac{1}{2}\chi'', \tag{24}$$

with $\chi \equiv \chi' + i\chi''$ being the complex susceptibility of the gain medium, and $|E|$ is the amplitude of the laser field. The measured linewidth, $\gamma_{MEAS}$, of a laser depends, of course, on the measurement bandwidth, $\Gamma_M$, which is the inverse of the measurement time, $\tau_M$. It has been shown [42,43,44] that the value of $\gamma_{MEAS}$ is given simply by the geometric mean of $\gamma_{STL}$ and $\Gamma_M$, so that we can write:

$$\gamma_{MEAS} = \sqrt{\gamma_{STL} \cdot \Gamma_M} = \sqrt{\frac{\hbar\omega}{2P_{out}(n_g\tau_c)^2\tau_M}(1+|\alpha|^2)}, \quad (25)$$

which can also be written as:

$$\gamma_{MEAS} = \frac{1}{n_g\tau_c} \cdot \sqrt{\frac{\hbar\omega}{2P_{out}\tau_M}(1+|\alpha|^2)}, \quad (26)$$

We consider first the expected value of $|\alpha|$ for the SLL. From the definition of $|\alpha|$ and Eq. 24, it follows that:

$$|\alpha| = \left(\frac{\partial\chi'}{\partial|E|}\right)/\left(\frac{\partial\chi''}{\partial|E|}\right), \quad (27)$$

We recall that in Eq. 6 and Eq. 7, we had defined $I$ as being $|E|^2$. Thus, we can use these equations to determine the numerator and the denominator in Eq. 27:

$$\frac{\partial\chi''}{\partial|E|} = \frac{4G_e\xi_e\Gamma_e^3|E|}{[2\Gamma_e\xi_e|E|^2+\Gamma_e^2+4(\nu-\nu_0)^2]^2} + \frac{4G_i\xi_i\Gamma_i^3|E|}{[2\Gamma_i\xi_i|E|^2+\Gamma_i^2+4(\nu-\nu_0)^2]^2}, \quad (28)$$

$$\frac{\partial\chi'}{\partial|E|} = -\frac{8G_e\xi_e\Gamma_e^2(\nu-\nu_0)|E|}{[2\Gamma_e\xi_e|E|^2+\Gamma_e^2+4(\nu-\nu_0)^2]^2} - \frac{8G_i\xi_i\Gamma_i^2(\nu-\nu_0)|E|}{[2\Gamma_i\xi_i|E|^2+\Gamma_i^2+4(\nu-\nu_0)^2]^2}. \quad (29)$$

Since the SLL operates at a frequency that is very close to $\nu_0$, we have $\Gamma_e \gg |\nu - \nu_0|$. Furthermore, since the background gain profile is very broad, compared to the laser Rabi frequency, we can assume $\Gamma_e \gg \xi_e|E|$. Since $G_e$ is much less than unity (see Table 1), we can thus ignore the first terms in both Eq. 28 and Eq. 29. We then get:

$$|\alpha| = \frac{2|\nu-\nu_0|}{\Gamma_i}. \quad (30)$$

For a typical range of values of $|\nu - \nu_0|$ within the primary mode zone, $|\alpha| \ll 1$. Thus, this parameter has a negligible effect on the linewidth of the SLL. It should be noted that, in general, the effect of $|\alpha|$ is found to be dominant only in systems where the laser frequency is far away from resonances in the gain media (such as in a semi-conductor laser [41]), and is negligible in a typical gas laser [41,42] where the laser frequency is generally close to the resonance in the gain medium.

Consider next the effect of the group index. As can be seen from Eq. 26, the linewidth decreases with increasing value of $n_g$. Intuitively, this can be understood as follows. The propagation of a pulse is slowed down by a factor of $n_g$. Thus, the effective travel time for a pulse as it propagates in the cavity increases by a factor of $n_g$. Therefore, the effective cavity decay time is $\tau_{c,eff} = n_g \tau_c$.

As we have noted above, Eq. 26 is not expected to be valid for a superluminal laser, since the dispersion in that case is anomalous (as noted before Eq. 23). The extent to which Eq. 26 is valid for a very large value of $n_g$ has not yet been tested experimentally. Realization of an SLL, followed by a careful measurement of $\gamma_{MEAS}$, would be a definitive way of answering this question. If Eq. 26 turns out to be valid for large values of $n_g$, then the SLL could also become potentially very useful as a source of radiation with extremely high spectral purity.

## 6. Conclusions

In this paper, we have described a subluminal laser which is extremely stable against perturbations of the cavity length. It makes use of a composite gain spectrum consisting of a broad background along with a narrow peak, produced via Raman gain in a lambda-type three level system. The stability of the laser is enhanced by a factor given by the group index, which can be as high as $10^5$ for experimentally realizable parameters. We have also shown that the quantum noise and measurement bandwidth limited linewidth of such a laser is expected to be smaller by the same factor. We first presented an analysis where the gain profile is modeled as a superposition of two Lorentzian functions. We then presented a numerical study based on a physical scheme for realizing the composite gain profile. In this scheme, the broad gain is produced by a high pressure buffer-gas loaded cell of rubidium vapor, pumped along the $D_2$ transition. The narrow gain is produced by using a second rubidium vapor cell, where optical pumping is used to produce a population inversion among the two ground state hyperfine sublevels, and a Raman pump is used to produce the gain. We have shown close agreement between the idealized model and the explicit model. A super-stable subluminal laser of this type, with reduced linewidth, may find many applications.

## 7. Acknowledgment

This research was supported in part by AFOSR Grant # FA9550-10-1-0228, NSF IGERT Grant # DGE-0801685, and NASA Grant # NNM13AA60C.

## 8. Figures and Tables

**Figure 1:** (a) Schematic of the configuration for a subluminal laser and (b) Raman transitions.

**Figure 2:** Gain profile based on the analytical model for an idealized SLL: (a) overall gain spectrum and (b) Raman gain.

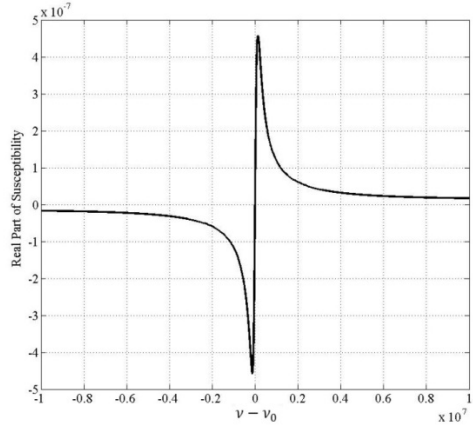

**Figure 3:** Real part of the susceptibility of the model for idealized SLL after lasing.

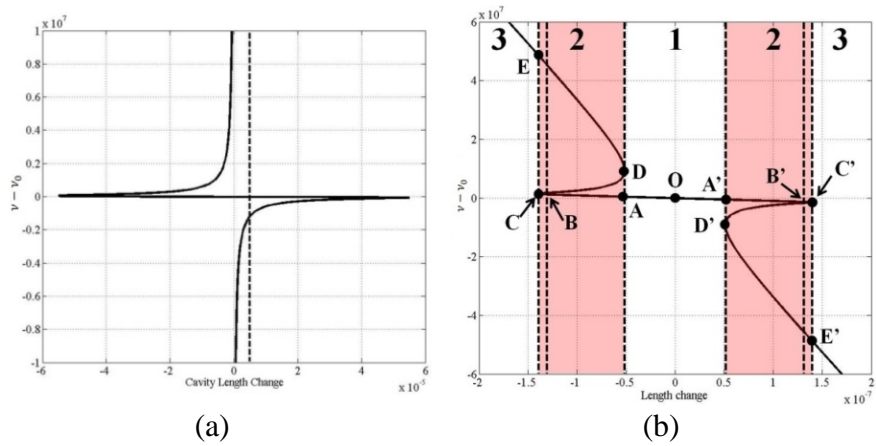

**Figure 4:** Relationship between cavity length and lasing detuning for the idealized SLL for (a) extremely high stability enhancement and (b) relatively low stability enhancement.

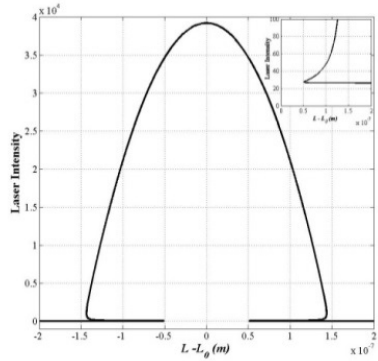 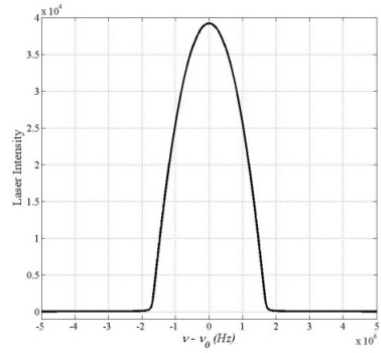

(a)                  (b)

**Figure 5:** Laser output intensity as a function of (a) length change and (b) laser frequency. The inset in Fig. 5(a) shows a blown up view of the multiple values of the intensity over a small vertical range.

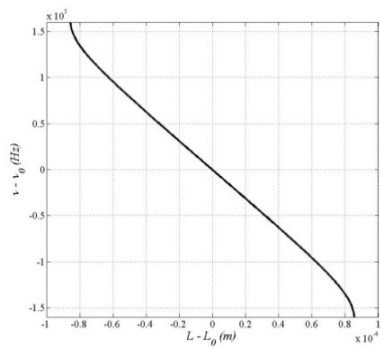 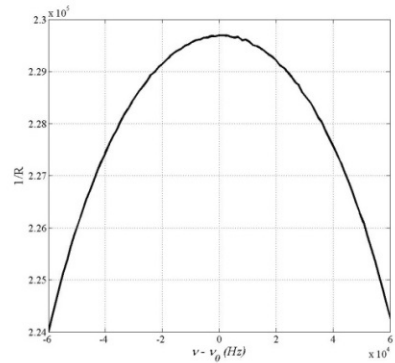

(a)                  (b)

**Figure 6:** Desired region of operation for the idealized SLL: (a) laser detuning vs. length change and (b) enhanced stability of the SLL.

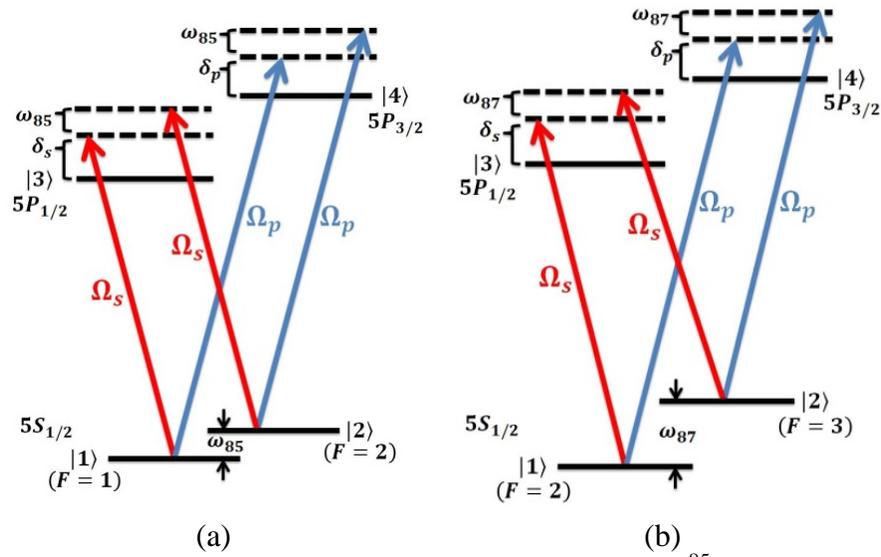

**Figure 7:** Transitions in the broad gain model for (a) $^{85}$Rb and (b) $^{87}$Rb.

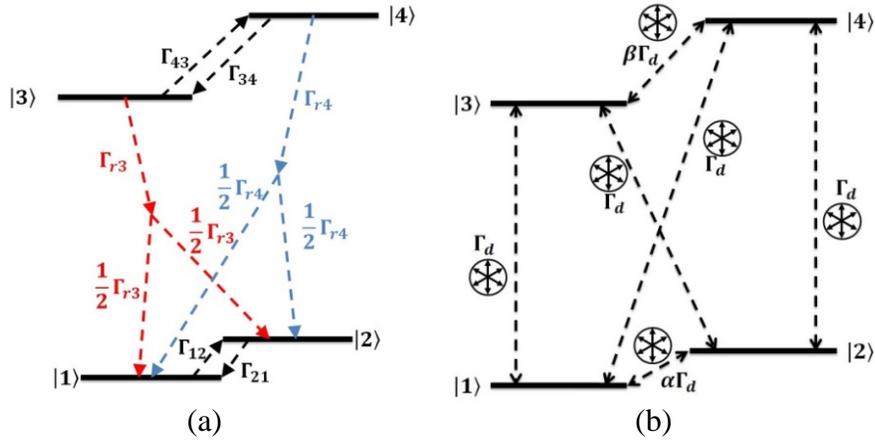

**Figure 8:** In the broad gain model: (a) population decay rates and (b) additional dephasing of coherence due to collisions.

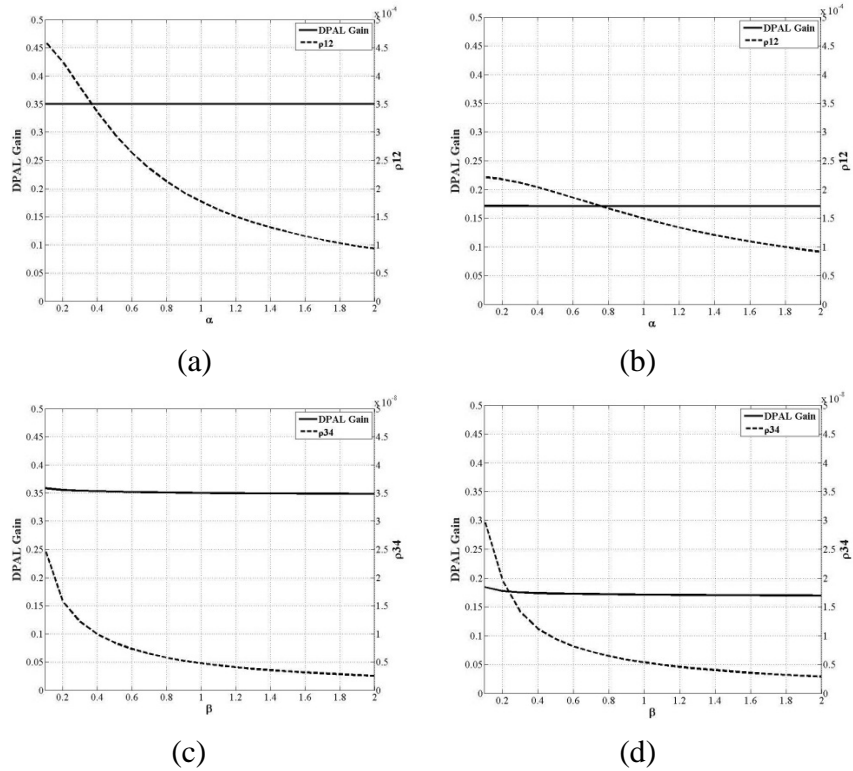

**Figure 9:** Maximum of broad gain versus parameters, as functions of: (a) $\alpha$ for $^{85}$Rb, (b) $\alpha$ for $^{87}$Rb, (c) $\beta$ for $^{85}$Rb, and (d) $\beta$ for $^{87}$Rb, which determine the BGI dephasing rates $\alpha\Gamma_d$ and $\beta\Gamma_d$.

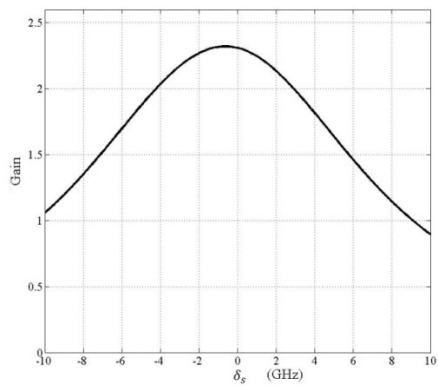

**Figure 10:** A typical spectrum of the gain medium in the DPAL with Ethane as the buffer gas.

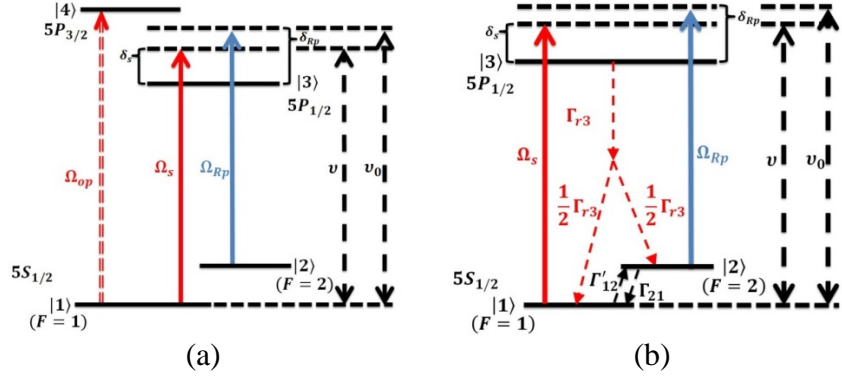

**Figure 11:** In the narrow gain model: (a) transitions, (b) effective 3-level model. Here $\Omega_{op}$, $\Omega_s$, and $\Omega_{Rp}$ are the Rabi frequencies for the optical pump, the signal field, and the Raman pump field, respectively. $\delta_s(\delta_{Rp})$ is the detuing of the signal (Raman pump) from the $|1\rangle - |3\rangle$ ($|2\rangle - |3\rangle$) transiton, $\nu$ is the frequency of the signal, and $\nu_0$ is defined as the value of the signal frequency corresponding to two photon resonance (i.e. $\delta_s = \delta_{Rp}$).

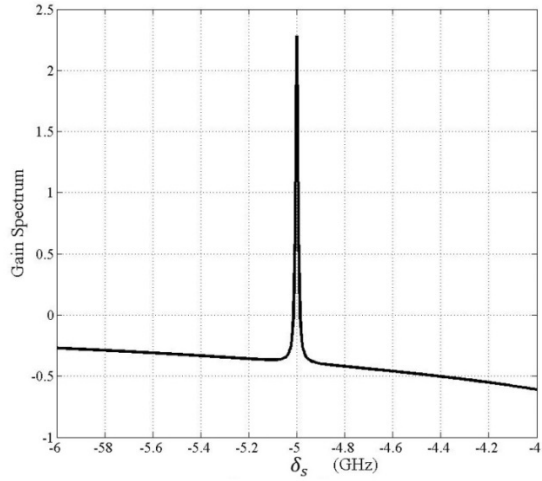

**Figure 12:** A typical gain spectrum produced by the Raman cell.

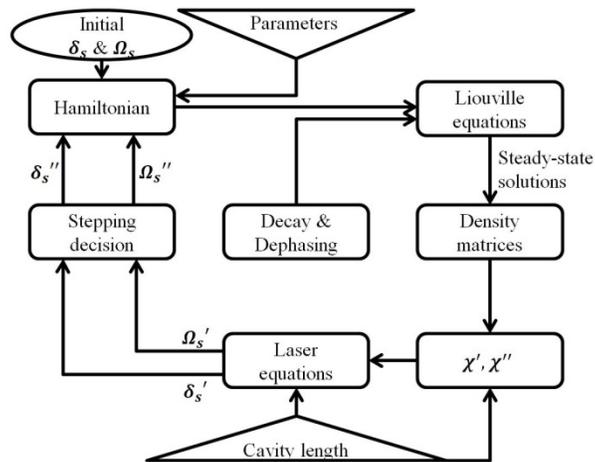

**Figure 13:** Flow-chart of the algorithm for solving the model for an actual SLL.

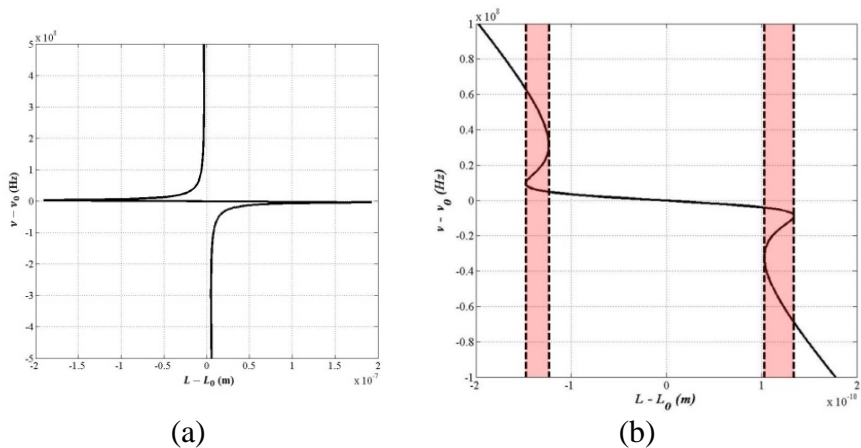

           (a)                                               (b)

**Figure 14:** Relationship between cavity length and lasing detuning for the practical SLL for (a) extremely high stability enhancement and (b) relatively low stability enhancement.

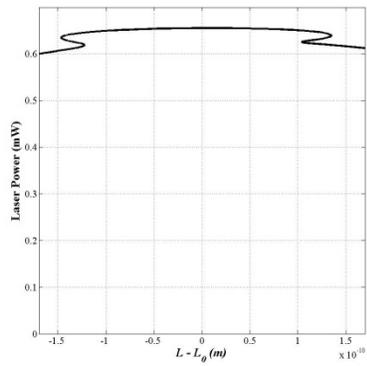 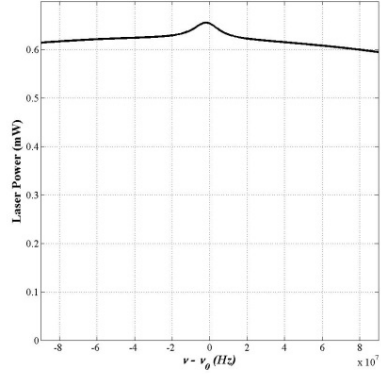

(a)                   (b)

**Figure 15:** Output power as a function of (a) length change and (b) laser frequency, for the model for a practical SLL.

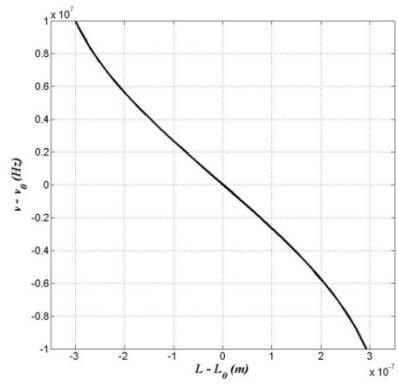 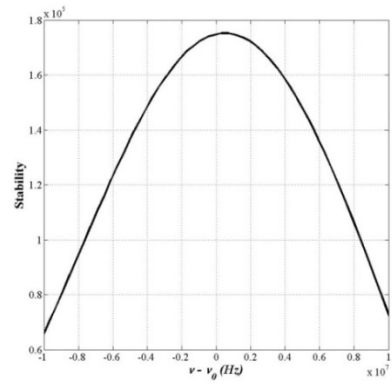

(a)                   (b)

**Figure 16:** Desired region of operation in the model for a practical SLL: (a) laser frequency as a function of cavity length and (b) stability as a function of laser frequency.

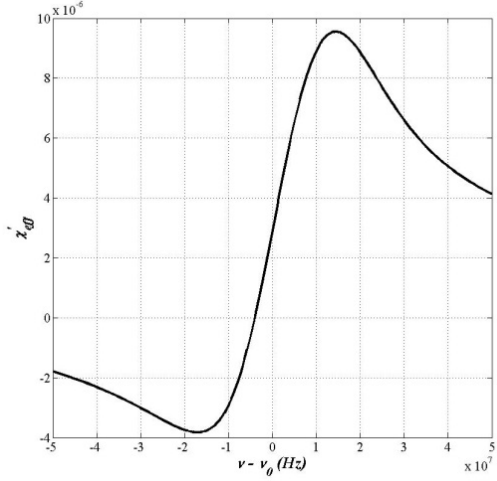

**Figure 17:** Real part of the effective dispersion in steady state of the model for a practical SLL.

**Caption List**

**Table 1** Parameters used in the analytical model.
**Table 2** Parameters used for generating gain spectrum of the DPAL.
**Table 3** Parameters used for generating gain spectrum of the medium in Raman cell.
**Table 4** Parameters used in the numerical model.
**Figure 1:** (a) Schematic of the configuration for a subluminal laser and (b) Raman transitions.
**Figure 2:** Gain profile based on the analytical model for an idealized SLL: (a) overall gain spectrum and (b) Raman gain.
**Figure 3:** Real part of the susceptibility of the model for idealized SLL after lasing.
**Figure 4:** Relationship between cavity length and lasing detuning for the idealized SLL for (a) extremely high stability enhancement and (b) relatively low stability enhancement.
**Figure 5:** Laser output intensity as a function of (a) length change and (b) laser frequency. The inset in Figure 5(a) shows a blown up view of the multiple values of the intensity over a small vertical range.
**Figure 6:** Desired region of operation for the idealized SLL: (a) laser detuning vs. length change and (b) enhanced stability of the SLL.

**Figure 7:** Transitions in the broad gain model for (a) $^{85}$Rb and (b) $^{87}$Rb.

**Figure 8:** In the broad gain model: (a) population decay rates and (b) additional dephasing of coherence due to collisions.

**Figure 9:** Maximum of broad gain versus parameters, as functions of: (a) α for $^{85}$Rb, (b) α for $^{87}$Rb, (c) β for $^{85}$Rb, and (d) β for $^{87}$Rb, which determine the BGI dephasing rates $\alpha\Gamma_d$ and $\beta\Gamma_d$.

**Figure 10:** A typical spectrum of the gain medium in the DPAL with Ethane as the buffer gas.

**Figure 11:** In the narrow gain model: (a) transitions, (b) effective 3-level model. Here $\Omega_{op}$, $\Omega_s$, and $\Omega_{Rp}$ are the Rabi frequencies for the optical pump, the signal field, and the Raman pump field, respectively. $\delta_s(\delta_{Rp})$ is the detuing of the signal (Raman pump) from the $|1\rangle - |3\rangle$ ($|2\rangle - |3\rangle$) transiton, $\nu$ is the frequency of the signal, and $\nu_0$ is defined as the value of the signal frequency corresponding to two photon resonance (i.e. $\delta_s = \delta_{Rp}$).

**Figure 12:** A typical gain spectrum produced by the Raman cell.

**Figure 13:** Flow-chart of the algorithm for solving the model for an actual SLL.

**Figure 14:** Relationship between cavity length and lasing detuning for the practical SLL for (a) extremely high stability enhancement and (b) relatively low stability enhancement.

**Figure 15:** Output power as a function of (a) length change and (b) laser frequency, for the model for a practical SLL.

**Figure 16:** Desired region of operation in the model for a practical SLL: (a) laser frequency as a function of cavity length and (b) stability as a function of laser frequency.

**Figure 17:** Real part of the effective dispersion in steady state of the model for a practical SLL.

---